\documentstyle[12pt,epsfig]{article}

\def\lsim{\mathrel{\rlap {\raise.5ex\hbox{$ < $}}
{\lower.5ex\hbox{$\sim$}}}}

\newcommand{\pr}{\paragraph{}}
\newcommand{\be}{\begin{equation}}
\newcommand{\ee}{\end{equation}}
\newcommand{\bea}{\begin{eqnarray}}
\newcommand{\nn}{\nonumber}
\newcommand{\eea}{\end{eqnarray}}
\newcommand{\nd}[1]{/\hspace{-0.6em} #1}

\newcommand{\Mpb}{\mbox{${\overline{p}}$}}
\newcommand{\kn}{\mbox{${K^{0}}$}}
\newcommand{\knb}{\mbox{${\overline{K}{}^{0}}$}}
\newcommand{\ko}{{K^{0}}}
\newcommand{\kob}{{\overline{K}{}^{0}}}
\newcommand{\ppb}{\mbox{${p\overline{p}}$}}
\newcommand{\kl}{\mbox{${K_{L}}$}}
\newcommand{\ks}{\mbox{${K_{S}}$}}
\newcommand{\km}{\mbox{${K^{-}}$}}
\newcommand{\kp}{\mbox{${K^{+}}$}}

\newcommand{\nnp}{\mbox{${\pi^{-}}$}}
\newcommand{\pp}{\mbox{${\pi^{+}}$}}

\newcommand{\mita}{\mbox{$|\eta_{+-}|$}}

\newcommand{\nora}{\mbox{$\alpha$}}
\newcommand{\mev}{\mbox{${MeV} \! / \! c$}}

\newcommand{\GeV}{\mbox{${GeV}$}}
\newcommand{\boldsymbol}{ }

\def\gappeq{\mathrel{\rlap {\raise.5ex\hbox{$>$}}
{\lower.5ex\hbox{$\sim$}}}}

\def\lappeq{\mathrel{\rlap{\raise.5ex\hbox{$<$}}
{\lower.5ex\hbox{$\sim$}}}}

\def\ie{{\em i.e.}}

\def\s{{\,\rm s}}
\def\S{S\hskip-8pt/\hskip2pt}
\def\H{H\hskip-8.5pt/\hskip2pt}

\def\beq{\begin{equation}}
\def\eeq{\end{equation}}

\def\I#1{{\rm Im}\,#1}
\def\Tr{{\rm Tr}\,}
\catcode`\@=11 
\def\coeff#1#2{{\textstyle{#1\over #2}}}

\def\ket#1{\left| #1\right\rangle}

\def\vev#1{\left\langle #1\right\rangle}
\def\lsim{\mathrel{\mathpalette\@versim<}}
\def\gsim{\mathrel{\mathpalette\@versim>}}
\def\@versim#1#2{\vcenter{\offinterlineskip
    \ialign{$\m@th#1\hfil##\hfil$\crcr#2\crcr\sim\crcr } }}

\def\JL{J. L. Lopez}
\def\DVN{D. V. Nanopoulos}

\def\t1{{\tilde 1}}

\def\to{\rightarrow}

\def\NPB#1#2#3{Nucl. Phys. B {\bf#1} (19#2) #3}
\def\PLB#1#2#3{Phys. Lett. B {\bf#1} (19#2) #3}

\def\PRD#1#2#3{Phys. Rev. D {\bf#1} (19#2) #3}

\def\gappeq{\mathrel{\rlap {\raise.5ex\hbox{$>$}}
{\lower.5ex\hbox{$\sim$}}}}

\def\lappeq{\mathrel{\rlap{\raise.5ex\hbox{$<$}}
{\lower.5ex\hbox{$\sim$}}}}

\begin{document}

\begin{titlepage}
\begin{flushright}
ENSLAPP-A-530/95\\
{\tt hep-ph/9506395}
\end{flushright}
\begin{center}
EXPERIMENTAL
TESTS OF CPT SYMMETRY AND QUANTUM MECHANICS
AT CPLEAR \\

\vglue 0.5cm

{\bf N. E. Mavromatos~$^{(a)~*}$ } \\
Laboratoire de Physique Th\`eorique
ENSLAPP (URA 14-36 du CNRS, associ\'ee \`a l' E.N.S
de Lyon, et au LAPP (IN2P3-CNRS) d'Annecy-le-Vieux),
Chemin de Bellevue, BP 110, F-74941 Annecy-le-Vieux
Cedex, France \\

  and \\

{\bf T. Ruf $^{*}$} \\
CERN PPE Division,
1211 Geneva 23, Switzerland \\

\vglue 0.3cm
\end{center}

\begin{abstract}
{\small
We review a phenomenological
parametrization of an open quantum-mechanical formalism for
CPT violation in the neutral kaon system,
and constrain the parameters using fits to recent CPLEAR data.
}
\end{abstract}

\vglue 0.2cm
\begin{center}
{\it Invited Contributions at the 2nd Da$\phi$ne
Workshop, DA$\phi$NE '95,
Frascati (Italy), April 4-8 1995 }

\vglue 0.1cm
$^{*}$~For the Collaboration:
John Ellis $^{(b)}$, Jorge L. Lopez $^{(c)}$,
N.~E.~Mavromatos, D.~V.~Nanopoulos$^{(c)}$,\\
and \\
the CPLEAR Collaboration~$^{\dagger}$
\end{center}

\vglue 0.2cm
\begin{flushleft}
{
R.~Adler$^2$, T.~Alhalel$^{2}$, A.~Angelopoulos$^1$, A.~Apostolakis$^1$,
E.~Aslanides$^{11}$, G.~Backenstoss$^2$, C.P.~Bee$^{2}$,
O.~Behnke$~^{17}$, A.~Benelli$~^{9}$,
V.~Bertin$^{11}$
% ...
}
\\ [1.2 cm]

{$^{(a)}$
On leave from P.P.A.R.C. Advanced Fellowship, Dept. of Physics
(Theoretical Physics), University of Oxford, 1 Keble Road,
Oxford OX1 3NP, U.K.  \\}
{$^{(b)}$CERN Theory Division,
1211 Geneva 23, Switzerland\\}
{$^{(c)}$Center for Theoretical Physics, Department of Physics, Texas
A\&M
University,
College Station, TX 77843--4242, USA,  and
Astroparticle Physics Group, Houston Advanced Research Center
(HARC),
The Mitchell Campus, The Woodlands, TX 77381, USA} \\
{
$^{\dagger}$The CPLEAR Collaboration:
$^{1}$University of Athens,
$^{2}$University of Basel,
$^{3}$Boston University,
$^{4}$CERN,
$^{5}$LIP and University of Coimbra,
$^{6}$Delft University of Technology,
$^{7}$Uni\-ver\-sity of Fribourg,
$^{8}$University of Ioannina,
$^{9}$Uni\-ver\-sity of Liverpool,
$^{10}$J. Stefan Inst. and Phys. Dep., University of Ljubljana,
$^{11}$CPPM, IN2P3-CNRS et Universit\'e d'Aix-Marseille II,
$^{12}$CSNSM, IN2P3-CNRS, Orsay,
$^{13}$Paul-Scherrer-Institut(PSI),
$^{14}$CEA, DSM/DAPNIA, CE-Saclay,
$^{15}$KTH Stockholm,
$^{16}$University of Thessaloniki,
$^{17}$ETH-IPP Z\"urich}
\end{flushleft}
\vspace{0.4in}

\end{titlepage}
\newpage
\baselineskip=14pt

\section{Introduction and Summary}
\pr
The neutral kaon system provides one of the most
sensitive laboratories for testing
quantum mechanics at the microscopic level
and for testing discrete symmetries
\cite{bellsteinberger}. It is the only place
where an experimental violation of CP
has yet been seen \cite{cronin}, and
it provides the strongest constraint on CPT
violation, via an upper bound on the difference
between the $\kn$ and the $\knb$
masses in the context  of quantum mechanics \cite{PDG}.
CPT symmetry is a property of quantum
field theory which follows from locality, causality
and Lorentz invariance \cite{luders}.
It is therefore particulary important
to look for CPT violation, which, if observed, would
require us to revise one or more of these
fundamental principles. In particular,
the possibility  of CPT violation
has been raised in the context
of quantum gravity \cite{wald}, as a result of a possible
modification of conventional quantum
field theory.

A framework for analyzing this possibility is provided
by the formulation \cite{EHNS} of open
quantum-mechanical systems which are
coupled to an unobserved environment.
This would
induce a loss of quantum coherence in the observed system,
which should be described by a density matrix
$\rho $ that obeys a modified quantum
Liouville equation
\be
{\dot \rho } = i [ \rho, H ] + \nd{\delta H} \rho
\label{one}
\ee
where the extra term $\propto \nd{\delta H}$ may be conjectured to arise
from quantum-gravitational effects and have a magnitude which is at
most ${\cal O}(m_K^2 / M_{Pl} )$ , where $M_{Pl} = 1.2  \times 10^{19}\,\GeV$
is the gravitational mass scale obtained from Newton's constant:
$M_{Pl} =G_N^{-\frac{1}{2}} $.
An equation of the form (\ref{one}) is supported by one
interpretation
of string theory \cite{emnqm},
but could have more general applicability.

In the case of the neutral kaon system, the open-system equation (\ref{one})
introduces \cite{EHNS} three CPT-violating parameters
$\alpha, \beta, \gamma $ if energy
and strangeness conservation are assumed,
in addition to the CPT-violating
parameters $\delta m = m_{\ko} - m_{\kob} $
and $\delta \Gamma = \Gamma _{\ko}
- \Gamma _{\kob} $
that can
be discussed
in the conventional quantum-mechanical framework. Here we
use recent CPLEAR data on tagged kaon decays\cite{Guyot,Ruf}
into $2\pi$ final states, together with information on \mita\ and
the
semileptonic
decay asymmetry, to obtain bounds on these CPT-violating parameters.
A more detailed account appears elsewhere \cite{elmn,elmncplear}.

\section{Formalism and relevant observables}
\label{sec:observables}
In this section we first
review aspects of the modifications (\ref{one}) of
quantum mechanics believed to be induced by quantum
gravity \cite{EHNS}, as argued
specifically in the
context of a non-critical string analysis \cite{emnqm}.
This provides a specific
form for the modification (\ref{one}) of the quantum
Liouville equation for the temporal evolution of the density matrix of
observable matter \cite{emnqm}
\begin{equation}
   \frac{\partial}{\partial t} \rho = i [ \rho, H] + {\delta\H} \rho
\qquad ; \qquad \delta\H \equiv i {\dot g}^i G_{ij} [g^j, \rho ]
\label{oneb}
\end{equation}
where the the $g^i$ are generic field-theory
couplings on the string world sheet, and $G_{ij}$
is a metric on the space of these couplings. The
extra term $\delta\H$ in (\ref{oneb}) is such that
the time evolution has
the following basic properties:
\begin{description}
\item (i) The total probability is {\it conserved} in time
\begin{equation}
       \frac{\partial}{\partial t} \Tr \rho = 0
\label{2.1}
\end{equation}
\item (ii) The energy $E$ is {\it conserved on the average}
\begin{equation}
  \frac{\partial}{\partial t} \Tr(E \rho ) = 0
\label{2.2}
\end{equation}
as a result of the {\it renormalizability} of the world-sheet $\sigma$-model
describing string propagation in a string space-time foam background.
\item (iii) The von Neumann entropy $S \equiv -k_B \Tr \rho \ln\rho $
increases {\it monotonically with time}
\begin{equation}
    \frac{\partial}{\partial t} S \ge 0
\label{2.3}
\end{equation}
which vanishes only if one restricts one's attention to critical (conformal)
strings, in which case there is no arrow of time \cite{emnqm}.
However,
we argue that quantum fluctuations in the
background space
time should be treated by including
non-critical (Liouville) strings \cite{aben,DDK}, in
which case (\ref{2.3}) becomes a strict
inequality. This latter property also
implies that the statistical entropy $S_{\rm st} \equiv \Tr \rho ^2 $ is also
monotonically increasing with time, pure states evolve into mixed ones and
there is an arrow of time in this picture \cite{emnqm}.
\item (iv) Correspondingly, the superscattering matrix $\S$, which is
defined by its action on asymptotic density matrices
\begin{equation}
  \rho _{out} = \S \rho _{in}
\label{2.4}
\end{equation}
cannot be factorised into the usual product of the Heisenberg scattering matrix
and its hermitian conjugate
\begin{equation}
     \S \ne S S^\dagger \qquad ; \qquad S=e^{-iHt}
\label{2.5}
\end{equation}
with $H$ the Hamiltonian operator of the system. In particular this property
implies that $\S$ has no inverse, which is also expected from the property
(iii).
\item (v) The absence of an inverse for $\S$ implies that {\it strong} CPT
invariance of the low-energy subsystem is lost, according
to the general analysis of \cite{wald,emnqm}.
\end{description}

It should be stressed that, although for the purposes of the present work we
keep the microscopic origin of the quantum-mechanics-violating terms
unspecified, it is only in the non-critical string model of Ref.~\cite{emnqm} -
and the associated approach to the nature of time - that a concrete microscopic
 model guaranteeing the properties (i)-(v) has so far emerged naturally.
Moreover, it is worth pointing out that
within the non-critical-string
framework, we expect
that the string $\sigma$-model coordinates $g^i$ obey
renormalization-group equations of the general form
\begin{equation}
    {\dot g}^i = \beta ^i M_{Pl} \qquad : \qquad
    |\beta ^i | = {\cal O}\left(\frac{E^2}{M_{Pl}^2}\right)
\label{E}
\end{equation}
where the dot denotes differentiation
with respect to the target time, measured in string $(M_{Pl}^{-1})$
units, and
$E$ is a typical energy scale in the observable
matter system. Since $G_{ij}$ and
$g^i$ are themselves dimensionless numbers of order unity,
we expect that
\begin{equation}
      |\delta\H | = {\cal O}\left(\frac{E^2}{M_{Pl}}\right)
\label{F}
\end{equation}
in general. However, it should be emphasized that there
are expected to be system-dependent numerical factors
that depend on the underlying string model, and that
$|\delta\H |$ might be suppressed by further
($E/M_{Pl}$)-dependent factors, or even vanish.
Nevertheless, (\ref{F}) gives us an order of magnitude
to aim for in the neutral kaon system, namely
${\cal O}((\Lambda_{\rm QCD}~{\rm or}~m_s)^2/M_{Pl}) \sim 10^{-19}$\,\GeV.

In the formalism of Ref.~\cite{EHNS}, the extra (non-Hamiltonian) term in the
Liouville equation for $\rho$ can be parametrized by a $4\times 4$ matrix
$\delta\H_{\alpha\beta} $, where the indices $\alpha, \beta, \dots$ enumerate
the Hermitian $\sigma$-matrices $\sigma _{0,1,2,3}$, which we represent in the
$K_{1,2}$ basis. We refer the reader to the literature \cite{EHNS,emncpt}
for details of this description, noting here the following
forms for the neutral kaon Hamiltonian
\begin{equation}
  H = \left( \begin{array}{cc}
  M - \coeff{i}{2}\Gamma - {\rm Re} M_{12} + \coeff{i}{2} {\rm Re} \Gamma _{12}
&  \coeff{1}{2}\delta m - \coeff{i}{4} \delta \Gamma
  -i {\rm Im} M_{12}  - \coeff{1}{2} {\rm Im} \Gamma _{12}  \\
  \coeff{1}{2}\delta m - \coeff{i}{4} \delta \Gamma
  + i {\rm Im} M_{12}  - \coeff{1}{2} {\rm Im} \Gamma _{12} &
  M - \coeff{i}{2}\Gamma + {\rm Re} M_{12} - \coeff{i}{2} {\rm Re} \Gamma _{12}
 \end{array}\right)
\label{nkham}
\end{equation}
in the $K_{1,2}$ basis, or
\begin{equation}
 H_{\alpha\beta}
 =\left( \begin{array}{cccc}  - \Gamma & -\coeff{1}{2}\delta \Gamma
& -{\rm Im} \Gamma _{12} & -{\rm Re}\Gamma _{12} \\
 - \coeff{1}{2}\delta \Gamma
  & -\Gamma & - 2{\rm Re}M_{12}&  -2{\rm Im} M_{12} \\
 - {\rm Im} \Gamma_{12} &  2{\rm Re}M_{12} & -\Gamma & -\delta m    \\
 -{\rm Re}\Gamma _{12} & -2{\rm Im} M_{12} & \delta m   & -\Gamma
\end{array}\right)
\label{hnk12}
\end{equation}
in the $\sigma$-matrix basis. As discussed in Ref.~\cite{EHNS}, we
assume that the dominant violations of quantum mechanics conserve strangeness,
so that $\delta\H_{1\beta }$ = 0, and that $\delta\H_{0\beta }$ = 0 so as to
conserve probability. Since $\delta\H_{\alpha\beta }$ is a symmetric
matrix, it follows that also $\delta\H_{\alpha 0}=\delta\H_{\alpha 1}=0$.
Thus, we arrive at the general parametrization
 \begin{equation}
  {\delta\H}_{\alpha\beta} =\left( \begin{array}{cccc}
 0  &  0 & 0 & 0 \\
 0  &  0 & 0 & 0 \\
 0  &  0 & -2\alpha  & -2\beta \\
 0  &  0 & -2\beta & -2\gamma \end{array}\right)
\label{nine}
\end{equation}
where, as a result of the positivity of the hermitian density matrix $\rho$
\cite{EHNS}
\begin{equation}
\alpha, \gamma  > 0,\qquad \alpha\gamma>\beta^2\ .
\label{positivity}
\end{equation}

We recall \cite{emncpt} that the CPT transformation can be expressed as
a linear combination of $ \sigma _{2,3}$ in the $K_{1,2}$ basis :
${\rm CPT} = \sigma_3 \cos\theta + \sigma_2 \sin\theta$
for some choice of phase $\theta$. It is
apparent that none of the non-zero terms $\propto   \alpha ,  \beta ,
 \gamma $ in $\delta\H_{\alpha\beta}$  (\ref{nine})
commutes with the CPT transformation. In other words, each of the three
parameters $\alpha$, $\beta$, $\gamma$ violates CPT, leading to a
richer phenomenology than in conventional quantum mechanics. This is
because the symmetric $\delta\H$ matrix has three parameters in its
bottom right-hand $2\times 2$ submatrix, whereas the $H$ matrix
appearing in the time evolution within quantum mechanics \cite{peccei}
has only one complex CPT-violating parameter $\delta$,
\begin{equation}
\delta = -\coeff{1}{2}
\frac{\coeff{1}{2}\delta\Gamma+i\delta m}{\coeff{1}{2}|\Delta\Gamma|+i\Delta m}
\label{cptdelta}
\end{equation}
where $\delta m$ and $\delta \Gamma $ violate CPT, but do not induce any mixing
in the time evolution of pure state vectors\cite{peccei,emncpt}.
The parameters
$\Delta m = M_L -M_S$ and $|\Delta\Gamma|=\Gamma_S-\Gamma_L$ are the
usual differences
between mass and decay widths, respectively, of $K_L$ and
$K_S$ states.
For more details we refer the reader to the literature
\cite{emncpt}. The above results imply that the experimental constraints
\cite{PDG} on CPT violation have to be rethought. There
are essential differences between
quantum-mechanical CPT violation\cite{peccei}
and
the non-quantum-mechanical CPT violation
induced by the effective parameters
$\alpha, \beta, \gamma$ \cite{EHNS,emncpt}.

Useful observables are
associated with
the decays of neutral kaons to $2\pi$ or $3\pi$ final states,
or semileptonic decays to $\pi l \nu$. In the density-matrix formalism
introduced above, their values are given \cite{EHNS} by expressions
of the form
\be
     < O_i > = Tr( O_i \rho )
\label{13and1/2}
\ee
where the observables $O_i$ are represented by $2 \times 2$ hermitian
matrices.
For future use, we give their
expressions in the $K_{1,2}$ basis
\begin{eqnarray}
 O_{2\pi} &=& \left( \begin{array}{cc} 0 & 0 \\
0 & 1 \end{array} \right)\ ,\qquad  O_{3\pi} \propto
\left( \begin{array}{cc} 1 & 0 \\
0 & 0 \end{array} \right)\ , \label{2pi-obs} \\
O_{\pi^-l^+\nu} &=& \left( \begin{array}{cc}
1 & 1 \\1 & 1 \end{array} \right)\ ,\qquad
O_{\pi^+l^- \bar\nu} = \left( \begin{array}{rr}
1 & -1 \\
-1 & 1 \end{array} \right)\ .
\label{semi-obs}
\end{eqnarray}
which constitute a complete hermitian set.
In this formalism,
pure $\kn$ or $\knb$ states, such as the ones used as
initial conditions in the CPLEAR experiment \cite{Guyot,Ruf}
are described by the following density matrices
\begin{equation}
\rho_{\ko} =\coeff{1}{2}\left( \begin{array}{cc}
1 &1 \\1 & 1 \end{array} \right)\ , \qquad
\rho_{\kob} =\coeff{1}{2}\left( \begin{array}{rr}
1 & -1 \\-1 & 1 \end{array} \right)\ .
\label{rhos}
\end{equation}
We note the similarity of the above density matrices (\ref{rhos})
to the semileptonic decay observables in (\ref{semi-obs}), which is
due to the strange quark ($s$) content of the kaon $\kn \ni {\bar s}
\rightarrow {\bar u} l^+ {\bar\nu} , \knb
 \ni s \rightarrow  u l^- \nu$, and our assumption of the validity of the
$\Delta S = \Delta Q$ rule.

Below, we shall apply the above formalism to compute\cite{elmn}
 the time evolution
of certain experimentally-observed quantities that are of relevance to the
CPLEAR experiment\cite{Guyot,Ruf}.
These are asymmetries associated with decays of an initial
$\kn$ beam as compared to corresponding decays of an initial $\knb$ beam
\begin{equation}
    A (t) = \frac{
    R(\knb_{t=0} \rightarrow
{\bar f} ) -
    R(\kn_{t=0} \rightarrow
f ) }
{ R(\knb_{t=0} \rightarrow
{\bar f} ) +
    R(\kn_{t=0} \rightarrow
f ) }\ ,
\label{asym}
\end{equation}
where $R(\kn\rightarrow f)\equiv \Tr[O_{f}\rho (t)]$, denotes the decay rate
into the final state $f$, given that one starts from a pure $ \kn$ at $t=0$,
whose density matrix is given in (\ref{rhos}), and
$R(\knb \rightarrow {\bar f}) \equiv \Tr [O_{\bar f} {\bar \rho}(t)]$
denotes the decay rate into the conjugate state ${\bar f}$, given that one
starts from a pure $\knb$ at $t=0$.

To determine the temporal evolution of the above observables, which is
crucial for experimental fits, it is necessary to know the equations of motion
for the components of $\rho$ in the $K_{1,2}$ basis. These are
\cite{EHNS,emncpt}\footnote{Since we neglect $\epsilon'$ effects and assume the
validity of the $\Delta S=\Delta Q$ rule, in what follows we also consistently
neglect ${\rm Im}\,\Gamma_{12}$ \cite{elmn}.}
\begin{eqnarray}
\dot\rho_{11}&=&-\Gamma_L\rho_{11}+\gamma\rho_{22}
-2{\rm Re}\,[({\rm Im}M_{12}-i\beta)\rho_{12}]\,,\label{rho11}\\
\dot\rho_{12}&=&-(\Gamma+i\Delta m)\rho_{12}
-2i\alpha\I{\rho_{12}}+({\rm Im}M_{12}-i\beta)(\rho_{11}-\rho_{22})\,,
\label{rho12}\\
\dot\rho_{22}&=&-\Gamma_S\rho_{22}+\gamma\rho_{11}
+2{\rm Re}\,[({\rm Im}M_{12}-i\beta)\rho_{12}]\,,   \label{rho22}
\end{eqnarray}
where $\rho$ represents either $\Delta\rho$ or $\Sigma\rho$, defined by
the initial conditions
\begin{equation}
\Delta\rho(0)=\left(\begin{array}{cc}0&-1\\-1&0\end{array}\right)\ ,\qquad
\Sigma\rho(0)=\left(\begin{array}{cc}1&0\\0&1\end{array}\right)\ .
\label{InitialConditions}
\end{equation}
corresponding to the difference between (sum of) initially-pure
$\kn$ and $\knb$ states.
In these equations, $\Gamma_L=(5.17\times10^{-8}\s)^{-1}$ and
$\Gamma_S=(0.8922\times10^{-10}\s)^{-1}$ are the inverse $K_L$ and $K_S$
lifetimes, $\Gamma\equiv(\Gamma_S+\Gamma_L)/2$, $|\Delta\Gamma| \equiv
\Gamma_S-\Gamma_L =(7.364 \pm 0.016 )\times 10^{-15}\,\GeV$, and $\Delta
m=530.0\times10^{7}\s^{-1}=3.489\times10^{-15}\,\GeV$ is the
$K_L-K_S$
mass difference. Also, the CP impurity parameter $\epsilon$ is given by
\begin{equation}
  \epsilon =\frac{{\rm Im} M_{12}}{\coeff{1}{2}|\Delta\Gamma|+i\Delta m }\ ,
\label{epstext}
\end{equation}
which leads to the relations
\begin{equation}
\I{M_{12}}=\coeff{1}{2}{|\Delta\Gamma||\epsilon|\over\cos\phi},\quad
\epsilon=|\epsilon|e^{-i\phi} \quad : \quad
\tan\phi={\Delta m\over {1\over2}|\Delta\Gamma|},
\end{equation}
with $|\epsilon|\approx2.2\times10^{-3}$ and $\phi\approx45^\circ$ the
``superweak" phase.

These equations are to be compared with the corresponding quantum-mechanical
equations of Ref.~\cite{peccei,emncpt}. The
parameters ${\delta}m$ and $\beta$ play similar roles, although they appear
with different relative signs in different places, because
of the symmetry of $\delta\H$ as opposed to the antisymmetry of the
quantum-mechanical evolution matrix $H$. These differences are important for
the asymptotic limits of the density matrix, and its impurity.
A pure
state will remain pure as long as $\Tr\rho^2=(\Tr\rho)^2$ \cite{EHNS}. In the
case of $2\times2$ matrices $\Tr\rho^2=(\Tr\rho)^2-2\det\rho$, and therefore
the purity condition is equivalently expressed as $\det\rho=0$.
The existence of the $\delta\H$ term (\ref{nine})
obviously violates this condition, thereby leading to
mixed states.

To make a consistent phenomenological study of the various quantities discussed
above, it is essential to solve the coupled system of equations (\ref{rho11})
to (\ref{rho22}) for intermediate times. This requires approximations in order
to get analytic results \cite{Lopez}.
Writing
\begin{equation}
 \rho_{ij}(t)=\rho^{(0)}_{ij}(t)+\rho^{(1)}_{ij}(t)+
\rho^{(2)}_{ij}(t)+\cdots
\label{perturb}
\end{equation}
where $\rho^{(n)}_{ij}(t)$ is proportional to
$\widehat\alpha^{p_\alpha}\widehat\beta^{p_\beta}
\widehat\gamma^{p_\gamma}|\epsilon|^{p_\epsilon}$, with
$p_\alpha+p_\beta+p_\gamma+p_\epsilon=n$, one can obtain a set of
differential
equations at each order in perturbation theory.

At order $n$ the differential equations are of the form
\begin{equation}
{d\over dt}
\left[e^{At}\rho^{(n)}_{ij}(t)\right]=
e^{At}\sum_{i'j'}\,\!\!'\rho^{(n-1)}_{i'j'}(t)
\end{equation}
which can be integrated straightforwardly in terms of the known
functions at the $(n-1)$-th order, and the initial condition
$\rho^{(n)}_{ij}(0)=0$, for $n\ge1$, \ie,
\begin{equation}
\rho^{(n)}_{ij}(t)=e^{-At}\int_0^t dt'\,
e^{At'}\,\sum_{i'j'}\,\!\!'\rho^{(n-1)}_{i'j'}(t')\ .
\label{solution}
\end{equation}
Following this straightforward (but tedious) procedure we can
obtain the
expressions for $\Delta\rho$ at any desired order.

\section{Analytical results}
\label{sec:analytical}
We now proceed to give explicit expressions for the temporal evolution of the
asymmetries $A_{2\pi}$, $A_{3\pi}$,
$A_{\rm T}$, $A_{\rm CPT}$, and $A_{\Delta m}$
that are the possible
objects of experimental study, in particular
by the CPLEAR collaboration
\cite{Guyot,Ruf}.  A more detailed account of these
results is given in ref. \cite{elmn}\footnote{See also
ref. \cite{HP}.}.

\subsection{$\boldsymbol{A_{2\pi}}$}
The formula for this asymmetry, as obtained by applying
the formalism of section 2, assumes the form
\begin{equation}
A_{2\pi} = \frac{
\Tr[O_{2\pi} {\bar \rho} (t)] -
\Tr[O_{2\pi} \rho (t)]}
{
\Tr[O_{2\pi} {\bar \rho} (t)] +
\Tr[O_{2\pi} \rho (t)]}
\equiv
\frac{\Tr[O_{2\pi} \Delta \rho (t)]}{\Tr[O_{2\pi} \sum \rho (t)]}\ ,
\label{a2pi}
\end{equation}
where the observable $O_{2\pi}$ is given in (\ref{semi-obs}), and
the $\Delta \rho$ and $\sum \rho $ density matrix elements are
given as above (\ref{InitialConditions}).
The
result for the asymmetry, to second order in the small parameters, can
be written most concisely as
\begin{eqnarray}
A_{2\pi}(t)&=&\Biggl\{
2|\epsilon|\cos\phi+4\widehat\beta\sin\phi\cos\phi
-8\widehat\alpha\sin\phi\cos\phi
(|\epsilon|\sin\phi-2\widehat\beta\cos^2\phi)
\nonumber\\
&&-2\sqrt{|\epsilon|^2+4\widehat\beta^2\cos^2\phi}
\,\,e^{{1\over2}(\Gamma_S-\Gamma_L)t}\,
\left[\cos(\Delta m
t-\phi-\delta\phi)+{2\widehat\alpha\over\tan\phi}
X_\alpha\right]\Biggr\}\nonumber\\
&&/\left\{1+e^{(\Gamma_S-\Gamma_L)t}\,[\widehat\gamma+|\epsilon|^2
-4\widehat\beta^2\cos^2\phi-4\widehat\beta|\epsilon|\sin\phi
]\right\}
\label{CPTx2}
\end{eqnarray}
where
${\widehat\alpha},{\widehat\beta},{\widehat\gamma}$
are scaled variables $a/\Delta \Gamma $,$\beta /\Delta
\Gamma $, $\gamma / \Delta \Gamma $, and
$X_\alpha \equiv \cos\delta \phi\sin (\Delta m t - \phi )
-\frac{1}{2} |\Delta \Gamma | t \tan\phi \cos(\Delta m t - \phi -
\delta \phi ) + \sin\phi\cos (\Delta m t - 2\phi - \delta \phi )$.

The above expression should be compared with
the usual
case (\ie, $\widehat\alpha=\widehat\beta=\widehat\gamma=0$)
\begin{equation}
A_{2\pi}(t)= {2|\epsilon|\cos\phi
-2|\epsilon|\,e^{{1\over2}(\Gamma_S-\Gamma_L)t}\,\cos(\Delta m t-\phi)\over
1+e^{(\Gamma_S-\Gamma_L)t}\,|\epsilon|^2}
\label{A2pi-usual}
\end{equation}
One can readily see whether CP violation can in fact vanish,
its effects mimicked by non-quantum-mechanical CPT violation. Setting
$|\epsilon|=0$ one needs to reproduce the interference pattern and also the
denominator. To reproduce the overall coefficient of the interference pattern
requires $2\widehat\beta\cos\phi=\pm|\epsilon|$. The denominator (neglecting
$\widehat\gamma$) becomes $-4\widehat\beta^2\cos^2\phi\to-|\epsilon|^2$
and we have the wrong sign. Another problem is that
$\delta\phi\to-{\rm sgn}(\widehat\beta){\pi\over2}$ and the interference
pattern is shifted significantly.
This means that the effects seen in the neutral
kaon system, and conventionally interpreted as  CP violation,
indeed cannot be due to the CPT violation.

\subsection{$\boldsymbol{A_{3\pi}}$}
Analogously, the formula for the $3\pi$ asymmetry,
to first order in the small parameters, is given by
\begin{equation}
    A_{3\pi}(t) = {
\left[2|\epsilon|\cos\phi-4\widehat\beta\sin\phi\cos\phi\right]
-2e^{-{1\over2}(\Gamma_S-\Gamma_L)t}
\left[{\rm Re}\eta_{3\pi}\cos\Delta mt - {\rm Im}\eta_{3\pi}\sin\Delta
 mt\right]\over
1+\widehat\gamma-\widehat\gamma e^{-(\Gamma_S-\Gamma_L)t}}\ ,
\label{3p2}
\end{equation}
with
\begin{equation}
{\rm
Re}\eta_{3\pi}=|\epsilon|\cos\phi-2\widehat\beta\sin\phi\cos\phi,\quad
{\rm Im}\eta_{3\pi}=|\epsilon|\sin\phi+2\widehat\beta\cos^2\phi\ .
\label{3p3}
\end{equation}
In the CPLEAR experiment, the time-dependent decay asymmetry into
 $\pi^0\pi^+\pi^-$ is measured~\cite{Guyot,Ruf},
 and the data is fit to obtain the best
values for ${\rm Re}\eta_{3\pi}$ and ${\rm Im}\eta_{3\pi}$.

\subsection{$\boldsymbol{A_{\rm T}}$}
In the CPT-violating case, to first order in the small
parameters, one finds
the following
time-dependent expression
for this asymmetry
\begin{eqnarray}
A_{\rm T}&=&{4|\epsilon|\over\cos\delta\phi}\nonumber\\
&&\left\{{e^{-\Gamma_L t}\cos(\phi-\delta\phi)+e^{-\Gamma_S
t}\cos(\phi+\delta\phi)
-2e^{-\Gamma t}\cos(\Delta m t-\delta\phi)\cos\phi
\over
e^{-\Gamma_L t}(1+2\widehat\gamma)+e^{-\Gamma_S t}(1-2\widehat\gamma)
-2e^{-\Gamma t}
[\cos\Delta m t+{2\widehat\alpha\over\tan\phi}(\sin\Delta m t
-\Delta m t\cos\Delta m t)]
}\right\}\nonumber\\
\label{At}
\end{eqnarray}
where
we have defined
\begin{equation}
\tan\delta\phi =-{2\widehat\beta\cos\phi\over|\epsilon|}\  + \dots,
\label{deltaphi}
\end{equation}
with the $\dots$ denoting higher order corrections\cite{elmn}.
The expression (\ref{At})
aymptotes to
\begin{equation}
A_{\rm T}\to
 {4|\epsilon|\cos(\phi-\delta\phi)\over\cos\delta\phi
 (1+2\widehat\gamma)}
={4|\epsilon|\cos\phi-8\widehat\beta\sin\phi\cos\phi\over1+2\widehat\gamma}\ .
\label{Ata}
\end{equation}

\subsection{$\boldsymbol{A_{\rm CPT}}$}
Following the discussion in section~\ref{sec:observables}, the formula for this
observable, as defined by the CPLEAR Collaboration \cite{Ruf}, is given by
Eq.~(\ref{asym}) with $f=\pi^-l^+\nu$ and $\bar f=\pi^+l^-\bar\nu$.
To first order, in both the CPT-conserving and CPT-violating cases, we find
\begin{equation}
A_{\rm CPT}=0\ .
\label{Acpt=0}
\end{equation}
We point out that this result is a quite distinctive signature of the
modifications of the quantum mechanics proposed in Ref.~\cite{EHNS,
emncpt}, since in the case of quantum-mechanical violation of CPT symmetry
\cite{peccei} there is a non-trivial change in $A_{\rm CPT}$, proportional to
the CPT-violating parameters $\delta m$ and $\delta\Gamma$. Indeed,
we obtain~\cite{elmn} the following first-order asymptotic result
\begin{equation}
A_{\rm CPT}^{\rm QM}\to4\sin\phi\cos\phi\,\widehat{\delta m}
+2\cos^2\phi\,\widehat{\delta\Gamma}\ ,
\label{acptqm}
\end{equation}
written in terms of the scaled variables. Part of the reason for this
difference is the different role played by $\delta m$ as compared to the
$\beta$ parameter in the formalism of Ref.~\cite{EHNS},
as discussed in detail
in Ref.~\cite{emncpt}. In particular, there are important sign
differences between the ways that $\delta m$ and $\beta$ appear in the two
formalisms, that cause the suppression to second order of any
quantum-mechanical-violating effects in $A_{\rm CPT}$, as opposed to the
conventional quantum mechanics case.

\subsection{$\boldsymbol{A_{\Delta m}}$}
Following Ref.~\cite{Guyot}, one can define $A_{\Delta m}$ as
\begin{equation}
A_{\Delta m}= {R(\kn\to\pi^+)+R(\knb\to\pi^-)-R(\knb\to\pi^+)
-R(\kn\to\pi^-)\over R(\kn\to\pi^+)+R(\knb\to\pi^-)+R(\knb\to\pi^+)
+R(\kn\to\pi^-)}
\label{Adeltmadef}
\end{equation}
in an obvious short-hand notation for the final states of the semileptonic
decays, where only the pion content is shown explicitly.  In the formalism
of section~\ref{sec:observables}, this expression becomes
in the quantum-mechanics-violating case
to first order
\begin{equation}
A_{\Delta m}=-{2e^{-\Gamma t}
\left[\cos\Delta m t+{2\widehat\alpha\over\tan\phi}(\sin\Delta mt-\Delta
mt
\cos\Delta m t)\right]\over
e^{-\Gamma_L t}(1+2\widehat\gamma)+e^{-\Gamma_S t}(1-2\widehat\gamma)}
\label{Adeltam2}
\end{equation}
Since $\widehat\gamma$ is negligible, this observable provides an {\em
exclusive} test of $\widehat\alpha$.

\section{Regeneration}
\label{sec:regeneration}

Regeneration involves the coherent scattering of a $\kn$ or $\knb$
off a nuclear target, which we assume can be described using the normal
framework of quantum field theory and quantum mechanics. Thus we
describe it by an effective Hamiltonian which takes the form
\begin{equation}
 \Delta H = \left( \begin{array}{cc}
 T + {\overline T}&  T - {\overline T}  \\
  T - {\overline T}& T + {\overline T}
    \end{array}\right)
\label{4.6.3}
\end{equation}
in the $K_{1,2}$ basis, where
\begin{equation}
T=\frac{2\pi N}{m_K}\, {\cal M}\,,\qquad
\overline T=\frac{2\pi N}{m_K}\,\overline{\cal M}
\label{4.6.2}
\end{equation}
with ${\cal M} = \vev{\kn|A|\kn}$ the forward $\kn$-nucleus scattering
amplitude
(and analogously for $\overline{\cal M}$), and $N$ is the nuclear
regenerator density.
The regenerator effects $\Delta H$
can in principle be included as a contribution to $H$ in the density
matrix equation:
\begin{equation}
   \partial _t \rho = -i [ H, \rho] + i \delta\H\rho
\label{4.6.4}
\end{equation}
where $\delta\H$ represents the possible CPT- and QM-violating term.
However we note that
the regenerator provides an `environment' that induces
an effective
CPT violation within quantum mechanics, due to the
inequivalent scattering of $\kn$ and $\knb$
by the regenerator material.

It may be adequate as a first approximation to treat the regenerator as
very thin, in which case we may use the impulse approximation,
and the regenerator changes  $\rho$ by an amount
\begin{equation}
  \delta \rho = -i[\Delta {\cal H}, \rho ]
\label{4.6.5}
\end{equation}
where
\begin{equation}
  \Delta {\cal H} = \int dt \Delta H
\label{4.6.6}
\end{equation}
Writing
\begin{equation}
\rho=\left(
\begin{array}{cc}\rho_{11}&\rho^*_{12}\\ \rho_{12}&\rho_{22}\end{array}
\right)\ ,
\end{equation}
in this approximation we obtain
\begin{equation}
\delta \rho = -i \Delta T
 \left( \begin{array}{cc}
2i{\rm Im}\rho_{12}&-\rho_{11}+\rho_{22}\\
\rho_{11}-\rho_{22}&-2i{\rm Im}\rho_{12}\end{array}\right)\ ,
\label{4.6.7}
\end{equation}
where
\begin{equation}
\Delta T \equiv \int dt (T - {\overline T} )\ .
\label{4.6.8}
\end{equation}
This change in $\rho $ enables the possible CPT- and QM-violating terms in
(\ref{4.6.4}) to be probed
in a new way, as we now discuss in a special case.

Consider
the idealization that the neutral $K$ beam is already in a
$K_L$ state :
\begin{equation}
  \rho = \rho _L \approx \left( \begin{array}{cc}
1&  \epsilon ^* + B^*  \\
  \epsilon + B &  |\epsilon |^2 + C
 \end{array}\right)
\label{4.6.9}
\end{equation}
where
\begin{equation}
B = -i2\widehat\beta\cos\phi\, e^{-i\phi}\qquad ; \qquad
C=\widehat\gamma-4\widehat\beta^2\cos^2\phi-4
\widehat\beta|\epsilon|\sin\phi
\label{4.6.10}
\end{equation}
Substituting Eqs.~(\ref{4.6.9},\ref{4.6.10}) into Eq.~(\ref{4.6.7}),
we find that in the  joint large-$t$ and impulse approximations
\begin{equation}
    \rho + \delta \rho =
  \left( \begin{array}{cc}
 1  + 2\Delta T {\rm Im}(\epsilon + B )
&  \epsilon ^* + B^*
  + i (1 - |\epsilon |^2 - C)\Delta T  \\
  \epsilon + B
  - i (1 - |\epsilon |^2 - C)\Delta T
&  |\epsilon |^2 + C  - 2\Delta T {\rm Im}(\epsilon + B)
 \end{array}\right)\ .
\label{4.6.11}
\end{equation}
We see that the usual semileptonic decay asymmetry observable
\begin{equation}
     O_{\pi^-l^+\nu} -
     O_{\pi^+l^-\bar\nu} = \left(\begin{array}{cc}
0 & 2 \\
2 & 0 \end{array}\right) \qquad ,
\label{4.6.12}
\end{equation}
which measures ${\rm Re}(\epsilon + B)$ in the case without the
regenerator, receives no contribution from the regenerator (\ie, $\Delta T$
cancels out in the sum of the off-diagonal elements). On the other hand, there
is a new contribution to the value of $R_{2\pi}=R(K_L\to2\pi)\propto{\rm
Tr}[O_{2\pi}\rho]=\rho_{22}$, namely
\begin{equation}
R_{2\pi} = |\epsilon|^2
+\widehat\gamma-4\widehat\beta^2\cos^2\phi-4
\widehat\beta|\epsilon|\sin\phi
-2\Delta T {\rm Im}(\epsilon + B)\ .
\label{4.6.15}
\end{equation}
The quantity ${\rm Im}(\epsilon + B)$ was not accessible directly to the
observable $R_{2\pi}$ in the absence of a regenerator.
Theoretically, the phases of $\epsilon$ and $B$ (\ref{4.6.10})
are fixed, \ie,
\begin{equation}
{\rm Im}(\epsilon +
B)=-|\epsilon|{\sin(\phi-\delta\phi)\over\cos\delta\phi}=
-|\epsilon|\sin\phi-2\widehat\beta\sin\phi\cos\phi\ .
\label{4.6.16}
\end{equation}
Nevertheless, this phase prediction should be checked, so the regenerator makes
a useful addition to the physics programme.

The above analysis is oversimplified, since the impulse approximation may not
be sufficiently precise, and the neutral $K$ beam is not exactly in a
$K_L$ state. However, it may serve to illustrate the physics interest of
measurements using a regenerator\cite{elmn}. We
note that measurements with different
thicknesses of regenerator should have a distinctive dependence $\Delta
R_{2\pi}\propto\Delta T$, which is a nice signature. Moreover, the cylindrical
geometry of the CPLEAR detector provides such different measurements ``for
free'' at different planar angles.

\section{CPLEAR Bounds on CPT-Violating Parameters}
\label{sec:Bounds}

\subsection{Description of the CPLEAR Experiment}
\begin{figure}\begin{center}
 \vspace{2cm}
 \parbox{10cm}{\epsfig{file=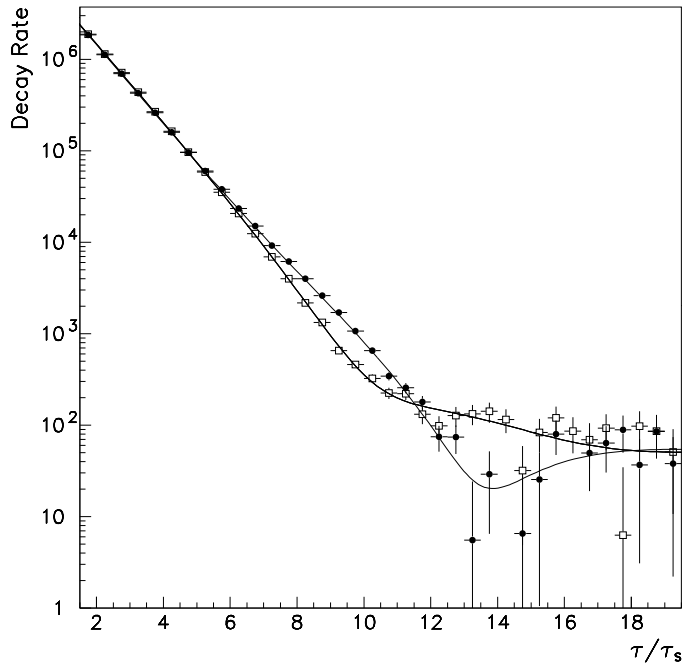,width=10cm}}
 \caption{Acceptance-corrected decay rate of \knb\ (filled circles) and
 \kn\ (open squares). Lines are the expected rates when the
Particle Data Group~$[3]$ values are used.}
 \label{fig:lifes}
 \end{center}  \end{figure}

 The CPLEAR experiment\cite{Guyot,noulis}
 is designed to determine CP and T violation
 in the neutral kaon system by measuring
 time-dependent decay rate asymmetries
 of CP and T conjugate processes. Initially-pure \kn\ and \knb\ states are
 produced concurrently in the annihilation channels
 $(\ppb)_{\hbox{\rm rest}}\rightarrow \kn \km\pp$ and
 $(\ppb)_{\hbox{\rm rest}}\rightarrow \knb \kp\nnp$, each one with a branching
 ratio of $\approx 0.2\%$. The strangeness of the neutral kaon is tagged by the
 charge sign of the accompanying kaon.
 In {\em Fig.~\ref{fig:lifes}} the decay rates for \kn\ and \knb\ measured by
 our experiment are shown separately, demonstrating the CP violation effect.

 A detailed description of the experiment can be found elsewhere
 \cite{cpex}, and only a few important items are mentioned here.
 The high rate of $200\,$\mev\ antiprotons ($10^6\,\Mpb/$s) is
 delivered by the LEAR machine at CERN. The antiprotons are stopped inside a
 gaseous hydrogen target of 16 bar pressure. A cylindrical detector is
 placed inside a solenoid of $1\,$m radius,
 $3.6\,$m length, providing a magnetic field of $0.44\,$T. The charged
 tracking system consists of two proportional chambers, six drift chambers and
 two layers of streamer tubes. Fast kaon identification is provided
 by a threshold \v{C}erenkov counter sandwiched between two scintillators.
 An electromagnetic calorimeter made of 18 layers of Pb converters
 and streamer tubes is used for photon detection and electron identification.
 An efficient and fast on-line data reduction is achieved with a multi-level
 trigger system based on custom-made hardwired processors.

 Kinematical constraints (energy-momentum conservation, \kn\ mass)
 and geometrical constraints (\kn\ flight direction and vertex separation)
 are used in the analysis to suppress the background from unwanted \kn\ decay
 channels and the background from \ppb\ annihilation events.
 In addition they improve the lifetime resolution.

 Knowing the initial strangeness of the neutral kaon, we are able to calculate
 time-dependent CP-violating decay-rate asymmetries:
 \begin{equation}
 A_{+-}(t)=
\frac{N_{{\overline{K}{}^{0}}\rightarrow \pi^+\pi^-}(t) - \nora
                    N_{{K^{0}}\rightarrow \pi^+\pi^-}(t)}
     {N_{{\overline{K}{}^{0}}\rightarrow \pi^+\pi^-}(t) + \nora
                    N_{{K^{0}}\rightarrow \pi^+\pi^-}(t)}
 \label {eq1}
 \end{equation}
 All the acceptances common for
 \kn\ and \knb\ cancel, thus reducing systematic uncertainties.
 The normalization factor $\nora$ is proportional to the tagging efficiency of
 \knb\ relative to \kn and is determined experimentally from the
 data together with the CP-violation parameters.

\subsection{Description of the fit}
 The results presented here are preliminary, and based on the CPLEAR data
 collected up to middle of 1994. The analysis based on standard quantum
 mechanics is presented in \cite{noulis}.

 The measured rates $N_{\kob\rightarrow \pi^+\pi^-}(t)$
 and $N_{\ko\rightarrow \pi^+\pi^-}(t)$ have been corrected for regeneration.
 The residual background contribution, mainly from semileptonic decays,
 is extracted from a Monte~Carlo simulation and taken into account in the fit.
 A systematic study of the effect of
 lifetime resolution and the regeneration correction is in preparation.

\begin{figure}\begin{center}
\parbox{10cm}{\epsfig{file=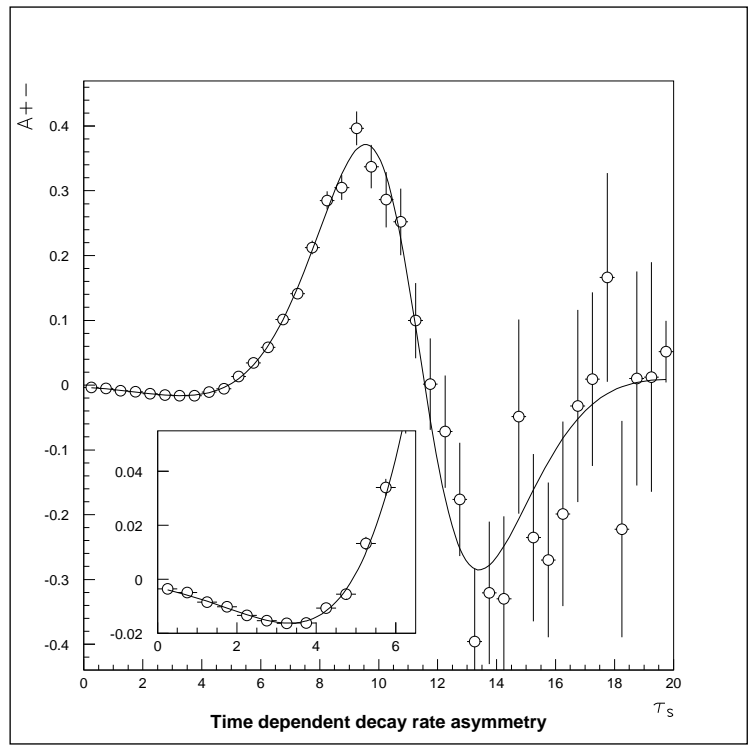,width=10cm}}
\caption{Decay-rate asymmetry as function of the decay eigentime.
 The line shows the result of the best fit. }
\label{fig:asym}
\end{center}  \end{figure}

We use the formalism for  $ \kn\rightarrow\pi^+\pi^-,
 \knb\rightarrow\pi^+\pi^-$ decays presented in section 3.1.
 In addition we require the following constraints to be fulfilled:
 \begin{itemize}
 \item $\vert\eta_{+-}\vert^2 = \hat{\gamma} ^2
 +\vert\varepsilon\vert^2
  {\displaystyle
   \frac{\cos(\varphi-2\delta\varphi)}{\cos(\varphi)
   \cos^2(\delta\varphi)}}$
  ,where $\vert\eta_{+-}\vert$ is the CP-violation parameter measured in \kl\
   decays to $\pi^+\pi^-$, taken from the Particle
Data Group~\cite{PDG},
   and $\tan(\delta\varphi)
  =   {\displaystyle -\frac{2\hat{\beta}\cos(\varphi)}{\varepsilon} }$,
 \item $\delta_L = 2\varepsilon\cdot\cos(\varphi)-4\hat{\beta}\cos(\varphi)
   \sin(\varphi)$,
   where $\delta_L$ is the semileptonic asymmetry measured in \kl\ decays,
   taken from the Particle Data Group~\cite{PDG},
 \item $\Delta m = (530.02\pm 1.63)\cdot 10^7\hbar{s}^{-1}$, is the
   average of the experiments listed in ref.~\cite{delm}.
 \end{itemize}
 Fitting the CP asymmetry shown in fig.~\ref{fig:asym} together with
 the above constraints, we obtain the following result for the CP-
 and CPT-violation parameters:
 \bea
  \varepsilon &=& (2.257\pm0.011)\times 10^{-3} \nn \\
 \hat{\alpha} &=& (-4.1\pm 8.5) \times 10^{-3}  \nn \\
 \hat{\beta}  &=& (-1.4\pm 4.6) \times 10^{-5}  \nn \\
 \hat{\gamma} &=& (-0.3\pm 3.3) \times 10^{-7}
\label{arr}
 \eea
 The $\chi^2$ of the fit is $0.8/{dof}$.

 Limits for $\hat{\alpha}$, $\hat{\beta}$ and $\hat{\gamma}$ are calculated by
 taking into account only the physical region,
i.e.,
 by imposing
the positivity constraints
  $\hat{\alpha}$,
 $\hat{\gamma} > 0$ and $\hat{\alpha}\cdot\hat{\gamma}>\hat{\beta}^2$.
 Integrating over the
 other two parameters we derive the following 90\% confidence level
 upper limits:
\bea
 \hat{\alpha} &<&           4.2\times 10^{-3} \nn \\
 \hat{\vert\beta\vert} &<&  2.1\times 10^{-5} \nn \\
 \hat{\gamma} &<&           2.9\times 10^{-7}
\label{array2}
\eea

\subsection{Remark on the $\boldsymbol{K_S-K_L}$ mass difference}
 In all experiments which have measured $\Delta m$ so far, standard quantum
 mechanics has been assumed. The CPLEAR collaboration has
 started to make a common fit of the
 semileptonic asymmetry $A_{\Delta m}$ and the two-pion asymmetry $A_{+-}$,
 keeping $\Delta m$ as a free parameter. These
 results will be published soon.

\section{Theoretical Comment on two-particle decay correlations}
\label{sec:HP}
Alternative
interesting tests of quantum
mechanics and CPT symmetry can be devised by
exploiting initial-state correlations due to
the production of a pair of neutral kaons
in a pure quantum-mechanical state, e.g.,
via $e^+e^- \rightarrow \phi \rightarrow \kn\knb $.
In this case, the initial state may
be represented by \cite{lipkin}
\begin{equation}
\ket{{\bf k}~;~{\bf - k}} = \coeff{1}{\sqrt{2}}
\left[ \ket{\kn ({\bf k} )~;~\knb ({\bf - k})}
-\ket{\knb ({\bf k} )~;~\kn ({\bf - k})} \right ]
\label{flavour}
\end{equation}
At subsequent times $t= t_1$ for particle $1$ and $t=t_2$ for particle $2$, the
joint probability amplitude is given in conventional quantum mechanics by
\begin{equation}
\ket{{\bf k},~t_1~;~{\bf - k},~t_2} \equiv
e^{-i H ({\bf k} ) t_1} e^{-i H ({\bf - k} ) t_2}
\ket{{\bf k}~;~{\bf - k}}
\label{lipkin}
\end{equation}
Thus the temporal evolution of the two-particle state is completely determined
by the one-particle variables (OPV) contained in $H$.

Tests
of quantum mechanics and CPT symmetry in $\phi $ decays \cite{dafnehb}
have recently been discussed \cite{HP} in a conjectured
extension of the formalism of \cite{EHNS,emncpt}, in which
the density matrix of the two-particle system
was hypothesized to be described completely in terms
of such one-particle variables (OPV): $ H $ and
$(\alpha, \beta, \gamma )$. It was pointed out that this
OPV hypothesis had several striking consequences,
including apparent violations of energy conservation
and angular momentum.

As we have discussed above,
the only known theoretical framework in which
the EHNS equation has been derived is that of a non-critical
string approach to string theory, in which (i) energy is conserved
in the mean as a consequence of the renormalizability
of the world-sheet $\sigma$-model, but (ii) angular momentum
is not necessarily conserved, as this is not guaranteed
by renormalizability and is not conserved in some
toy backgrounds~\cite{emnroma},
though we cannot exclude the possibility that it may be conserved
in some particular backgrounds. Therefore,
we are not concerned
that \cite{HP} find angular momentum non-conservation in their
hypothesized OPV approach, but the absence of energy conservation
in their approach leads us to the conclusion
that irreducible two-particle parameters must be introduced
into the evolution of the two-particle
density matrix.  The appearance
of such non-local parameters does not concern
us, as the string is intrinsically non-local in target
space, and this fact plays a key role in
our model calculations of contributions to $\delta\H$.
The justification and parametrization of such irreducible
two-particle effects goes beyond the scope
of these talks, and we plan to study
this subject in more detail in due course.

\section{Conclusions}
\label{sec:conclusions}

We have
discussed in these talks approximate
expressions for a complete  set of neutral kaon
decay observables ($\pi\pi, 3\pi, \pi^{\pm}l^{\mp}\nu )$,
which can be used to constrain the
parameters $\alpha, \beta, \gamma $
characterising CPT violation
in a formalism motivated by ideas about quantum gravity
and string theory, that incorporates a possible microscopic
loss of quantum coherence by treating the neutral kaon
as an open quantum-mechanical system \cite{EHNS,emnqm,emncpt,elmn}.
Detailed fits to recent CPLEAR experimental data
on $2\pi$ decays have been reported based on
our formulae\cite{elmncplear}. These may be
used to obtain indicative upper bounds
\be
  |\alpha | \lsim  3.1\times 10^{-17}\,
  \GeV,~|\beta | \lsim 1.5\times 10^{-19}\, \GeV,~|\gamma | \lsim 2.1\times
 10^{-21}\,\GeV
\label{8.1}
\ee
which are comparable with the order of magnitude
$\sim 10^{-19}\, \GeV $ which theory indicates\cite{emnqm} might
be attained by such CPT- and quantum-mechanics-violating parameters.

We have not presented explicit expressions for the case where
the deviation $\epsilon '/\epsilon \lsim 10^{-3} $ from pure
superweak CP violation is non-negligible, but our methods
can easily be extended to this case. They can also be used to obtain
more complete expressions for experiments with a regenerator,
if desired. Details of
the extension of the formalism of ref. \cite{EHNS}
to
correlated
$\kn\knb$ systems
produced in $\phi $ decay, as at
DA$\phi$NE \cite{dafnehb}, involves the introduction
of two-particle variables, which
lies beyond the scope of this paper.

We close by reiterating that the neutral kaon system
is the best microscopic laboratory for testing
quantum mechanics and CPT symmetry. We believe
that violations of these two fundamental principles,
if present at all, are likely to be linked, and have proposed
a formalism that can be used to explore systematically
this hypothesis, which is motivated by ideas about
quantum gravity and string theory. Our understanding
of these difficult issues is so incomplete
that we cannot calculate the sensitivity which would
be required to reveal modifications of quantum
mechanics or a violation of CPT. Hence we cannot
promise success in any experimental search for such
phenomena. However, we believe that both the theoretical
and experimental communities should be open to their
possible appearance.

\section*{Acknowledgments}
We would like to thank P. Eberhard
for useful discussions.
The work of N.E.M. has been supported by a
European Union Research Fellowship,
Proposal Nr. ERB4001GT922259, and that of
D.V.N. has been supported in part by DOE grant DE-FG05-91-ER-40633.
N.E.M. thanks D. Cocolicchio, G. Pancheri, N. Paver
and other members of the DA$\phi$NE working groups for their
interest in this work.


\begin{thebibliography}{99}
\bibitem{bellsteinberger} J.S. Bell and J. Steinberger, proc. 1965 Oxford
Intern. Conf. on Elementary Particles (Oxford 1966) , p. 165.
\bibitem{cronin} J.H. Christenson, J.W. Cronin, V.L. Fitch and R. Turlay,
Phys. Rev. Lett. 13 (1964) , 138.
\bibitem{PDG} {\it Review of Particle Properties},
Particle Data Group, Phys. Rev. D50 (1994), 1173.
\bibitem{luders} G. L\"uders, Ann. Phys. (NY) 2 (1957), 1.
\bibitem{wald} R. Wald, Phys. Rev. D21 (1980), 2742 ;
\par D.N. Page, Gen. Rel. Grav. 14 (1982), 299 ;
\bibitem{EHNS} J. Ellis, J.S. Hagelin, D.V. Nanopoulos and
M. Srednicki, Nucl. Phys. B241 (1984), 381.
\bibitem{emnqm}
J. Ellis, N.E. Mavromatos
and D.V. Nanopoulos,
Phys. Lett. B293 (1992), 37;
\par J. Ellis, N.E. Mavromatos and D.V. Nanopoulos,
preprint CERN-TH.7195/94, ENS-LAPP-A-463/94, ACT-5/94, CTP-TAMU-13/94,
lectures presented (by J.E. and N.E.M.)
at the {\it Erice Summer School of Sub-Nuclear Physics,
31st Course: From
Supersymmetry
to the Origin of Space-Time}, Ettore Majorana Centre, Erice, July
4-12, 1993, hepth 9403133, to be published in the
proceedings;
\par For a pedagogical review of this approach see:
D.V. Nanopoulos, Riv. Nuov. Cim. Vol. 17, No. 10 (1994), 1.
\bibitem{Guyot} The CPLEAR Collaboration, in Proceedings of the
XXVI International
Conference on High Energy Physics,
ed. by J. R. Sanford (AIP
Conference Proceedings No. 272), p. 510.
\bibitem{Ruf} The CPLEAR collaboration,
{\it Measurement of the CP violation parameter
$\eta_{+-}$ using tagged \kn\ and \knb}, submitted to Phys.
Lett. B.
\bibitem{elmn} \par J. Ellis, J.L. Lopez, N.E. Mavromatos
and D.V. Nanopoulos,
CERN, ENS-LAPP and Texas A \& M Univ. preprint,
ACT-06/95,
CERN-TH.95-99, ENS-LAPP-A-521/95, CTP-TAMU-16/95 (1995), hep-ph 9505340.
\bibitem{elmncplear} The CPLEAR
Collaboration and
J.~Ellis, \JL, N.E. Mavromatos, D.V. Nanopoulos,
in preparation.
\bibitem{aben} I. Antoniadis, C. Bachas, J. Ellis
and D.V. Nanopoulos, Phys. Lett. {\bf B211} (1988), 393;
Nucl. Phys. {\bf B328} (1989), 117; Phys. Lett. {\bf B257} (1991), 278.
\bibitem{DDK}F. David, Mod. Phys. Lett. {\bf A3} (1988), 1651;
\par J. Distler and H. Kawai, Nucl. Phys. {\bf B321} (1989), 509;
\par N.E. Mavromatos and J.L. Miramontes,
Mod. Phys. Lett. {\bf A4} (1989), 1847.
\bibitem{emncpt}  J. Ellis, N. E. Mavromatos and D.V. Nanopoulos,
\PLB{293}{92}{142} and CERN-TH.6755/92.
\bibitem{peccei} C.D. Buchanan, R. Cousins, C. O. Dib,
R.D. Peccei and J. Quackenbush, \PRD{45}{92}{4088};
C.O. Dib and R.D. Peccei, \PRD{46}{92}{2265}.
\bibitem{Lopez} \JL\ in {\em Recent Advances in the Superworld}, Proceedings of
the HARC Workshop,  The Woodlands,
April 14-16, 1993, edited by \JL\ and \DVN\
(World Scientific, Singapore 1994), p.~272.
\bibitem{HP} P. Huet and M. Peskin, \NPB{434}{95}{3}.
\bibitem{lipkin} H. Lipkin, \PLB{219}{88}{474}.
\bibitem{noulis} P.~Pavlopoulos, these proceedings.
\bibitem{cpex}   The CPLEAR Collaboration, {\it
The CPLEAR detector at CERN}, in preparation.
\bibitem{delm}   Geweniger et al., Phys.Lett. B52(1974) 108; \\
                 Gjesdal et al., Phys.Lett. B52(1974) 113; \\
                 L.K.~Gibbons et al., Phys.Rev.Lett. 70(1993) 1199; \\
                 B.~Schwingenheuer et al., EFI 94-60, submitted
                 for publication. \\
                 The CPLEAR Collaboration, {\it
                 New measurement of the \ks-\kl\ mass
 difference by using semileptonic decays of tagged neutral kaons}, in
 preparation.
\bibitem{dafnehb} {\it Da$\phi$ne Physics Handbook}, edited by
L. Maiani, L. Pancheri and N. Paver (INFN, Frascati, 1992).
\bibitem{emnroma} J. Ellis, N. E. Mavromatos and
D.V. Nanopoulos, Proc. {\it
`First International Conference on Phenomenology of
Unification
from Present to Future'}, Rome, `La Sapienza' University,
March 22-26 1994, p. 187 (World Sci. ); preprint
CERN-TH.7269/94, CTP-TAMU 26/94,
ENSLAPP-A-474/94, hep-th/9405196.

\end{thebibliography}
\end{document}